# Zero-dispersion wavelength mapping in short single-mode optical fibers using parametric amplification

A. Mussot[(1)], E. Lantz[(1)], A. Durécu-Legrand[(2)],
C. Simonneau[(2)], D. Bayart[(2)], T. Sylvestre[(1)], H. Maillotte [(1)]

[(1)] Département d'Optique P.M. Duffieux, Institut FEMTO-ST, UMR CNRS/Université de Franche-Comté n°6174, 25030 Besançon cedex, France
[(2)] Alcatel, Research and Innovation Department, Route de Nozay, 91460 Marcoussis, France
**herve.maillotte@univ-fcomte.fr**

Abstract: We demonstrate a novel convenient nondestructive method based on optical parametric amplification that allows retrieval of the zero-dispersion wavelength map along a short optical fiber span with a high spatial resolution. The improved resolution relies on the high sensitivity to the local longitudinal dispersion fluctuations of the parametric high-gain spectrum.

*Indexing terms: fiber nonlinear optics, fiber optical parametric amplifier, fiber characterization, zero-dispersion wavelength*

Processing of optical fibers unavoidably induces longitudinal fluctuations of chromatic dispersion, in particular in their zero-dispersion wavelength (ZDW), due to random variations in opto-geometric parameters of the preform and drawing-induced fluctuations. These ZDW fluctuations can have a detrimental impact in fiber-based communication systems and devices such as dispersion-managed soliton transmissions, four-wave mixing (FWM) frequency converters or fiber optical parametric amplifiers (FOPA) [1-3]. Therefore there is an increasing demand in accurately mapping the ZDW of single-mode optical fibers, and especially short spans of new generation nonlinear or microstructure fibers. Most previous non-destructive measurement techniques use FWM conversion efficiency [4-6]. They have been applied to long fiber spans with a good accuracy but give a spatial resolution limited to about a hundred meters. In this paper, we propose a novel method which takes advantage of the high sensitivity of the parametric high-gain spectrum to local dispersion variations in a FOPA. By demonstrating this technique in a highly-nonlinear optical fiber (HNLF) whose length is only a few hundred meters, we clearly show its enhanced spatial resolution, in the meter range, and its good precision.

Near the ZDW, the parametric gain depends on the phase mismatch relation [7]:

$$\kappa = \frac{(2\pi c)^3}{\lambda_P^4} \beta_3 (\lambda_0 - \lambda_P)(\lambda_P - \lambda_S)^2 + \left(\frac{2\pi c}{\lambda_P^2}\right)^4 \beta_4 (\lambda_P - \lambda_S)^4 + 2\gamma P_0 \qquad (1)$$

with $\beta_{3,4}$ the 3$^{rd}$ and 4$^{th}$ order dispersion terms, $\lambda_{P,S}$ the pump and signal wavelengths, $\gamma$ the nonlinear coefficient, $P_0$ the pump power, and $\lambda_0$ the ZDW. In a FOPA, longitudinal variations in ZDW will sharply and locally modify the phase-matching condition between the pump, the signal and the idler. Thus the phase-sensitive parametric gain spectrum can be dramatically modified on a short propagation length [2,3]. High gain parametric amplification is thus expected to be more sensitive to the ZDW fluctuations along the fiber than FWM.

Thanks to these properties, our method uses a FOPA to measure several experimental gain spectra at different $\lambda_P$'s around the ZDW of the amplifying fiber. Then, the $\lambda_0$ variations along the fiber are mapped from these spectra by using a set of orthogonal polynomials (of successive orders, designed for polynomial regression [8]) whose coefficients are optimized with a Gauss-Newton iterative inversion algorithm. In this way, once the algorithm converges, the experimental gain spectra are accurately fitted from numerical simulations based on an extended nonlinear Schrödinger equation (NLSE) that includes all known experimental parameters and the retrieved ZDW map.

To assess the efficiency of the method, we first generate numerically several parametric gain curves that correspond to a given ZDW map and demonstrate we can retrieve this map. We use typical values for $\beta_3$ (or dispersion slope, $D_S$) and $\beta_4$ in the HNLF [2,7]. The ZDW map to be retrieved (solid line in Fig.1-(a)) is generated in a 300-m long HNLF by making a random summation of sine functions [2] that properly model the medium- and long- scale $\lambda_0$ fluctuations measured in the literature [4]. Note that we deliberately choose a strong variation ($\Delta\lambda_0 \sim 5$ nm) on this short fiber length in order to show the high spatial resolution of our method. Using this ZDW map, we then numerically integrate the extended NLSE, which accounts for the higher-order dispersion terms as well as for stimulated Raman scattering and attenuation. The results of our simulations are presented in Fig.1-(b) that shows the different gain spectra obtained by tuning $\lambda_P$ with respect to the mean ZDW $\overline{\lambda_0}$. As expected, the gain profile is extremely dependent on the pump wavelength [3]. Starting from another arbitrary initial ZDW distribution within the fiber, we then use the Gauss-Newton algorithm to adjust the coefficients of 14 orthogonal polynomials (up to the 13th order) so as to model the ZDW fluctuations which provide the best fitting of these gain curves. For example, it can be convenient to start from the zero-order polynomial giving a constant ZDW distribution $\overline{\lambda_0}$, which is a measurable quantity [9] (dashed line in Fig.1-(a)). Then, the successive increasing order of each polynomial accounts for a higher spatial frequency of the $\lambda_0$ fluctuation, the largest order thus giving the maximum resolution.

The mean-square error (mean squared difference between the original and the retrieved curves) obtained after convergence of the algorithm is $1.5 \times 10^{-8}$ dB$^2$, which means that the reconstructed gain curves are not distinguishable from the departure ones, nor are the retrieved and generated ZDW maps, on the scale of Fig.1. Actually, the ZDW map, the $D_S$ and $\beta_4$ values are all retrieved with a maximum error smaller than 0.1% (in percentage of the total variation $\Delta\lambda_0$ for $\lambda_0$). Therefore in this ideally noiseless case the method proves to be consistent and accurate, with a high spatial resolution.

The experimental set-up [3,10] is a FOPA operating in continuous-wave (cw)-pumping regime with a 490-m long HNLF provided by Sumitomo Electric, whose nominal parameters are $\alpha$=0.56 dB/km, $D_S$=0.032 ps/nm$^2$/km, $\gamma$=11.2 W$^{-1}$.km$^{-1}$, and $\overline{\lambda_0}$=1553 nm. $\beta_4$ is not known. Fig.2-(b) shows the measured gain spectra obtained by tuning the pump wavelength in a 2-nm range (symbols). Note that the gain profile exhibits strong variations, especially when $\lambda_P$ varies in some tenths of nanometers around $\overline{\lambda_0}$. This indicates that the ZDW fluctuations in this HNLF are expected to be weak and, hence, that this fiber should have a very good longitudinal homogeneity. Using the procedure described above, we have retrieved the ZDW map that leads to these gain spectra. The ZDW shift, shown in solid line in Fig.2-(a), presents variations with a maximum amplitude of 1.5 nm. The reconstructed gain spectra obtained from NLSE simulations are plotted in Fig.2-(b), which shows that the numerical results (solid lines) fit very well the measured ones. For this best adjustment the mean-square error is 0.64 dB$^2$. The algorithm gives also $\overline{\lambda_0}$=1553 nm and $D_S$ = 0.027 ps/nm$^2$/km, in close agreement with the nominal values, while the determined coefficient $\beta_4$ = -3.6x10$^{-56}$ s$^4$.m$^{-1}$ is typical for a HNLF [2,7]. Actually in the measurements, the spatial resolution is limited by dispersion averaging of the parametric gain spatial dynamics on several meters [1,2]. Hence, the retrieved ZDW map exhibits the $\lambda_0$ fluctuations that have already been identified to occur on medium to long correlation length scales (some tens to several hundred meters) [1,2,4,5].

To validate these results, we have compared our ZDW map with two other independent measurements on the same fiber span provided by the manufacturer using two different methods: a phase-shift technique [9], which gives an accurate measurement of the mean ZDW (horizontal dashed line in Fig.2-(a)), and a standard FWM method [4] mapping the ZDW as indicated by the dotted line in Fig.2-(a). Our ZDW map clearly lies between both other measurements apart for the initial section. Moreover, the mean ZDW value given by our method is exactly the same as that measured by the phase-shift technique. Figs.2-(c) and 2-(d) present the results of additional counter-simulations where the parametric gain spectra calculated from the NLSE, when including either of the two other ZDW maps, are compared with the experimental measurements. For these simulations we use the nominal dispersion slope while $\beta_4$ has been adjusted to yield a minimum mean-square error in both fittings. As Fig.2-(c) shows, the discrepancy between the calculated profiles and the measured data dramatically increases, leading to a minimum mean-square error of 5.5 dB$^2$, when keeping $\lambda_0$ constant at 1553 nm over the entire fiber span. The discrepancy is even more pronounced when using the map retrieved from the FWM method (Fig.2-(d)), with a minimum mean-square error of 60 dB$^2$. This means that the solid ZDW map of Fig.2-(a) allows the achievement of a much better fitting of the experimental gain spectra. Hence, this comparative study proves both the validity and higher accuracy of our method as well as its efficiency in a practical situation, especially when handling short fiber spans.

In practice, an optimal choice of the set of gain curves (their number and the pump wavelengths) as well as the ideal polynomial number is a trade-off between simulation time consumption and accuracy. As a rule, choosing gain spectra with $\lambda_P$ very close to $\overline{\lambda_0}$ yields an enhanced sensitivity to the ZDW fluctuations and in the determination of $\beta_4$, while those obtained at sufficiently higher wavelengths (*i.e.* in anomalous dispersion) yield better precision in the determination of $\overline{\lambda_0}$ and D$_S$. Moreover, attempting to retrieve too many polynomials could result in systematic divergences. Our experimental gain curves have been retrieved with 4 polynomials, *i.e.* 6 fitted coefficients including the determination of $D_S$ and $β_4$. On the other hand, for an adequate number of polynomials, the method converges always to the same solution for a range of initial parameters and clearly diverges outside this range. Hence the solution, if any, is unique.

Otherwise, if a good agreement between the experimental gain curves and the retrieved ones is a necessary condition to guarantee the accuracy of the retrieved ZDW map and dispersion coefficients, it is not a sufficient condition. Indeed, in the Gauss-Newton optimization method, huge uncertainties could occur if the matrix connecting the retrieved fiber parameters to the gain spectra is nearly singular because of similar effects of two parameters on the gain profile for example [11]. Fortunately, a standard technique involving the inversion of this matrix exists to calculate the uncertainty on the parameters [11]. This technique necessitates knowledge of the covariance of the noise on the gain curves. We have adapted it to the evaluation of the uncertainties on the ZDW map by using the following steps: (1) the uncertainty on a gain measurement is evaluated from the mean-square error and the number of independent gain measurements is roughly estimated as the number of intersections between the measured curves and the retrieved ones; (2) the covariance of the parameters is then estimated in a standard way by inverting the matrix connecting the parameters to the gain curves (this covariance gives directly the uncertainties on $D_S$ and $β_4$); (3) new uncorrelated parameters are computed by diagonalizing this covariance matrix; (4) 1000 ZDW maps are computed by adding to these uncorrelated parameters a random error with Gaussian probability and the standard deviation computed at the previous step; and (5) the uncertainty range at 95% of confidence on the ZDW map (between the solid lines on Fig.3) is estimated by eliminating at each fiber abscissa the 25 minimum and 25 maximum values. The results given on Fig.3 confirm the good accuracy of our method.

In conclusion, using a FOPA in the cw-pumping regime, we have demonstrated a novel convenient method to map accurately the zero-dispersion wavelength and determine dispersion slope and fourth-order dispersion coefficient of a short optical fiber span. By taking advantage of the strong sensitivity of the parametric high-gain spectra to local dispersion variations, the spatial resolution of ZDW mapping is enhanced, which makes the method relevant for short optical fibers like HNLF's or microstructure fibers. It can also be applied to longer spans of conventional fibers, provided that phase-matching can be fulfilled with a sufficiently long coherence length. For long spans however, polarization-mode dispersion and birefringence of the fiber could become limiting factors for the spatial resolution [5] and would require implementing more elaborated retrieval procedure.

We gratefully acknowledge Sumitomo Electric for providing the HNLF and the data of independent ZDW measurements.

**Figure captions:**

Fig. 1. (a) Numerically generated and retrieved ZDW map (solid line), and mean ZDW (dashed). (b) Parametric gain spectra for different pump wavelengths. Parameters of the NLSE simulation: $\alpha$=0.56 dB/km, $\overline{\lambda_0}$=1550 nm, $D_S$=0.032 ps/nm$^2$/km, $\beta_4$=-2x10$^{-55}$ s$^4$.m$^{-1}$, $\gamma$=11.2 W$^{-1}$.km$^{-1}$, L=300 m, $P_0$=600 mW.

Fig. 2. (a) ZDW map retrieved from our method (solid line), a FWM method (dotted), and mean ZDW measured by a phase-shift method (dashed). (b) Measured gain spectra at several pump wavelengths (symbols) and spectra numerically reconstructed from the solid ZDW map in (a) (solid lines). (c) and (d), same as (b) but with numerical gain spectra reconstructed from, respectively, the dashed (constant ZDW) and the dotted ZDW maps of (a). The continuous pump powers launched in the HNLF are 635 mW (circles and crosses), 645 mW (plus signs), and 620 mW (squares).

Fig. 3. Estimation of the accuracy: retrieved ZDW map (dashed), minimum and maximum uncertainties (dotted), with 95% of the events located between the solid lines. At ± 2 standard deviations corresponding to 95% of confidence, the uncertainties on $D_S$ and $\beta_4$ are: $D_S$=0.027 ± 0.003 ps/nm$^2$/km, $\beta_4$=-3.6 ± 0.6 10$^{-56}$ s$^4$.m$^{-1}$.

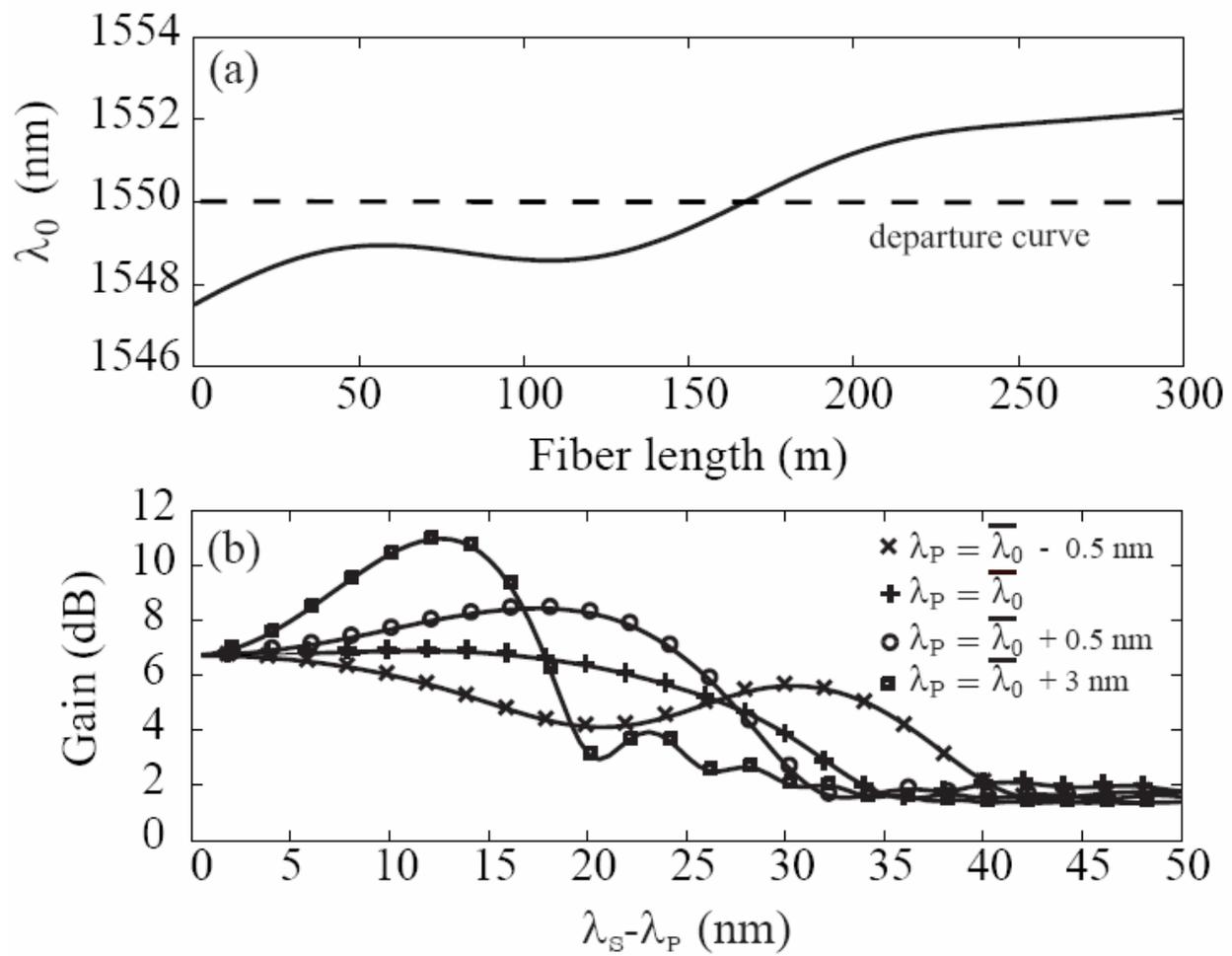

**Figure 1: A. Mussot *et al.*, "Zero-dispersion wavelength mapping ..."**

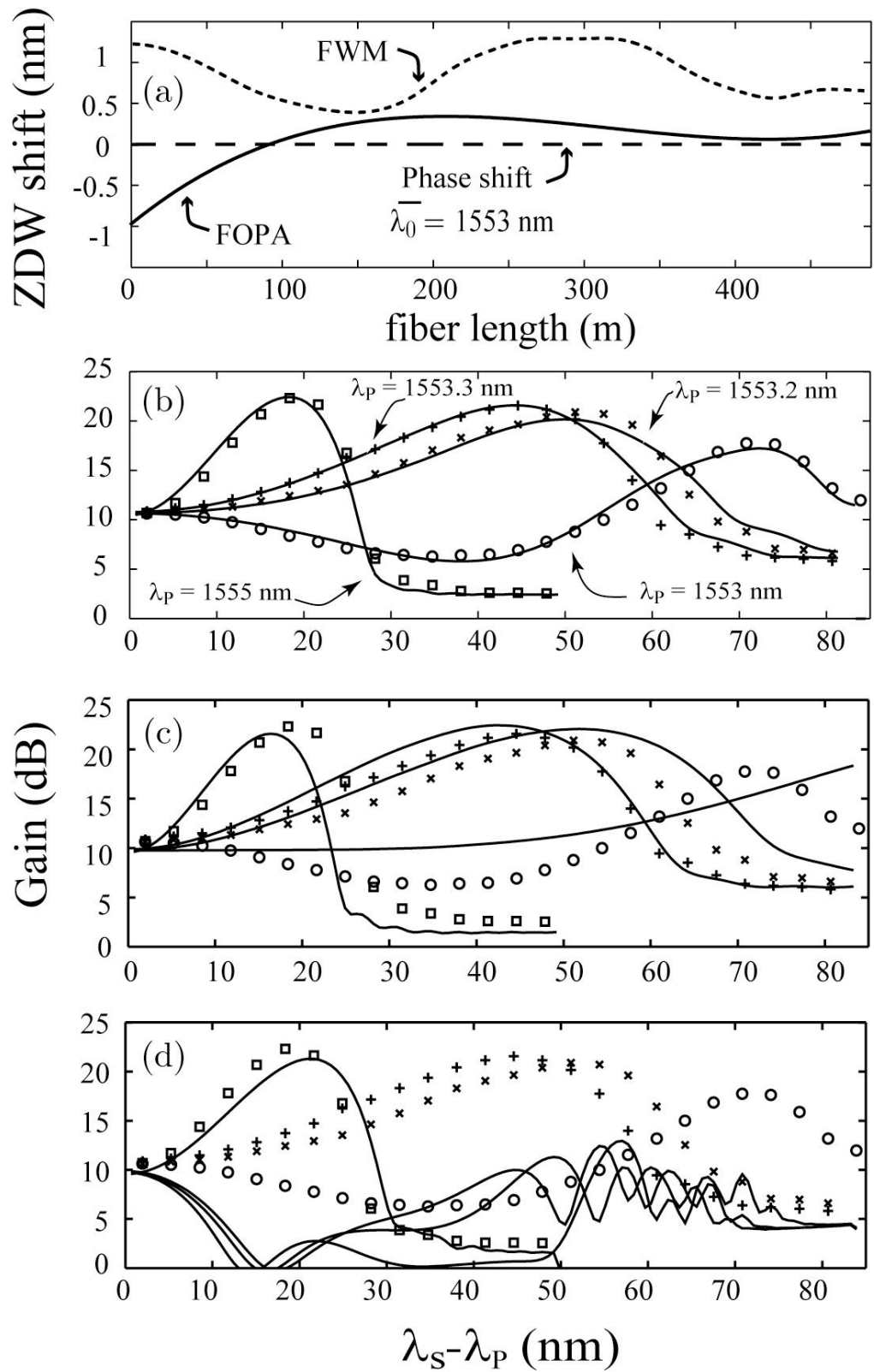

**Figure 2: A. Mussot *et al.*, "Zero-dispersion wavelength mapping ..."**

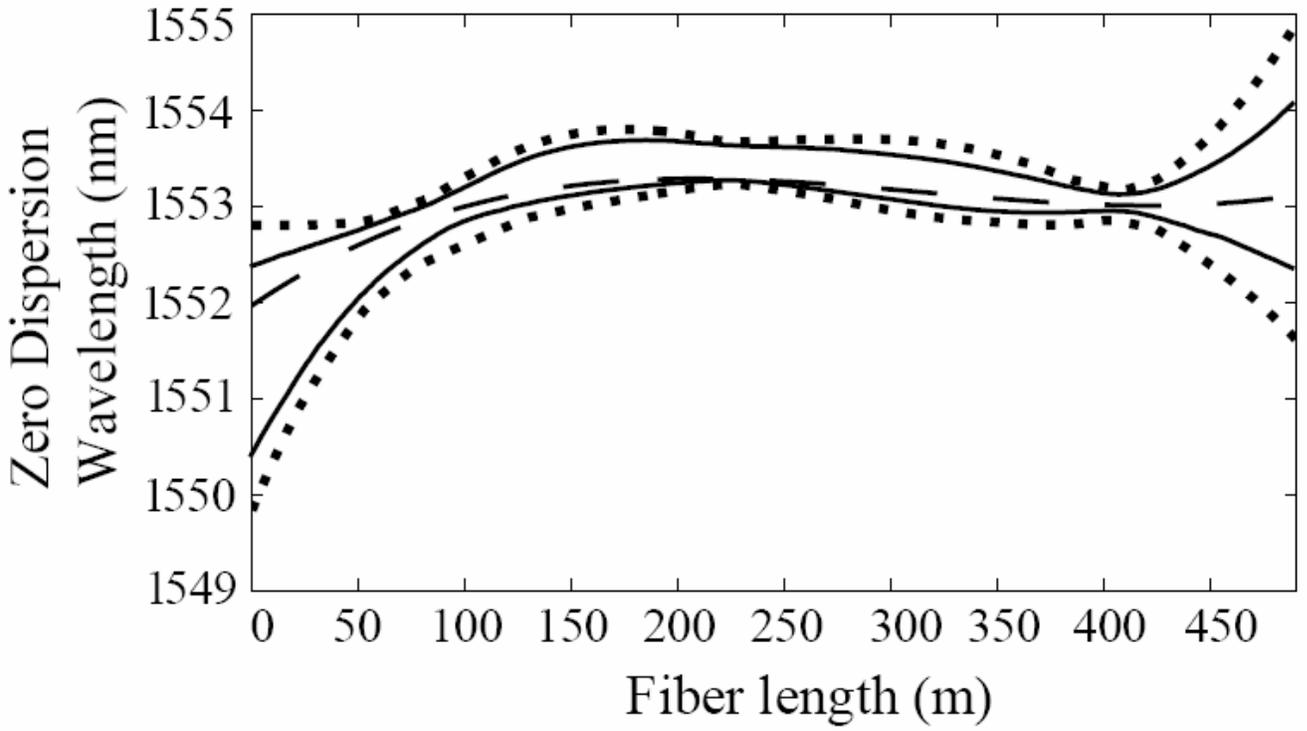

**Figure 3: A. Mussot *et al.*, "Zero-dispersion wavelength mapping ..."**